\documentclass{article}
\usepackage[T1]{fontenc}
\usepackage{geometry}
\usepackage{amsmath}

\makeatletter

\providecommand{\LyX}{L\kern-.1667em\lower.25em\hbox{Y}\kern-.125emX\@}

\usepackage{amsfonts}
\usepackage{amssymb}
\makeatother

\begin{document}

\title{On generalizations of Verlinde's formula}

\author{P. Bantay\\
Inst. for Theor. Phys., Eotvos Univ.}

\maketitle
\begin{abstract}
It is shown that traces of mapping classes of finite order may be expressed
by Verlinde-like formulae. The 3D topological argument is explained, and the
resulting trace identities for modular matrix elements are presented.
\end{abstract}
One of the basic results of 2D Conformal Field Theory is the celebrated formula
of Verlinde \cite{Verlinde}
\begin{equation}
\label{verlinde}
N_{pqr}=\sum _{s}\frac{S_{ps}S_{qs}S_{rs}}{S_{0s}}
\end{equation}
(as usual, we label by \( 0 \) the vacuum of the theory), expressing the fusion
rule coefficient \( N_{pqr} \) in terms of the matrix elements of the modular
transformation \( S:\tau \rightarrow \frac{-1}{\tau } \). Denoting by \( \mathcal{V}_{g}\left( p_{1},\ldots ,p_{n}\right)  \)
the space of genus \( g \) holomorphic \( n \)-point blocks with the insertions
\( p_{1},\ldots ,p_{n} \), and remembering that \( N_{pqr}:=\dim \mathcal{V}_{0}\left( p,q,r\right)  \),
Eq.(\ref{verlinde}) implies the more general result 
\begin{equation}
\label{verlinde2}
\dim \mathcal{V}_{g}\left( p_{1},\ldots ,p_{n}\right) =\sum _{q}S_{0q}^{2-2g}\prod _{i=1}^{n}\frac{S_{qp_{i}}}{S_{0q}}
\end{equation}
 Eqs.(\ref{verlinde}) and (\ref{verlinde2}) are not only of importance for
physics, but they have raised much attention in the mathematical literature
as well.

The virtue of Verlinde's formula is that it expresses the dimension of the space
\( \mathcal{V}_{g}\left( p_{1},\ldots ,p_{n}\right)  \) of holomorphic blocks,
or what is the same, the trace of the identity operator acting on \( \mathcal{V}_{g}\left( p_{1},\ldots ,p_{n}\right)  \),
in terms of seemingly unrelated quantities, matrix elements of modular transformations
acting on the space of genus 1 characters. But besides the identity, there are
other operators of interest that act naturally on \( \mathcal{V}_{g}\left( p_{1},\ldots ,p_{n}\right)  \):
the operators representing the transformations of the mapping class group \( M_{g,n} \)
of genus \( g \) surfaces with \( n \) punctures. It seems therefore reasonable
to ask whether there exists Verlinde-like expressions for the traces of mapping
classes other than the identity. This question has been addressed previously
in \cite{BV}, and it was found that there exists such expressions for some
mapping classes. The following problems have been left open by that work:

\begin{enumerate}
\item Characterise those mapping classes for which there exist a Verlinde-like formula
expressing the trace in terms of the matrix elements of modular transformations.
\item Give a simple recipe to write down the trace formula in case there is one.
\end{enumerate}
The purpose of the present note is to solve the above problems in the case of
closed surfaces, i.e. for \( n=0 \).

Before giving the answers and sketching the argument leading to them, let's
point out an important consequence already noticed in \cite{BV}. Namely, one
gets trace formulae already at genus \( 1 \), i.e. for modular transformations
acting on the space of genus 1 characters, resulting in non-trivial algebraic
identities for modular matrix elements, e.g. one has 
\begin{equation}
\label{traceS}
Tr\left( S\right) =\sum _{p}\frac{\left( S^{-1}T^{4}S\right) _{p0}\left( S^{-1}T^{4}S\right) _{p0}\left( S^{-1}T^{-2}S\right) _{p0}}{S_{p0}}
\end{equation}
which should be compared with the obvious answer 
\[
Tr\left( S\right) =\sum _{p}S_{pp}\]
The equality of the two expressions for \( Tr\left( S\right)  \) may be shown
to be independent of the usual consistency requirements on modular matrix elements,
i.e. Verlinde's theorem and the modular relation, and lends itself to a simple
numerical check. The relevant trace identities arising this way will be summarised
later in Table 1.

Let's now turn to the actual topic of this note, i.e. giving a simple characterisation
of those mapping classes whose trace may be expressed by a Verlinde-like formula.
As we shall argument later, there is such a trace formula in case the mapping
class has a fixed point in its action on Teichmuller-space. By a theorem of
Nielsen \cite{Hempel}, this is equivalent to the assertion that the mapping
class has finite order. In particular for \( g=1 \) this means that we get
trace formulae for the powers of \( S \) and \( ST \), leading to non-trivial
trace identities..

To give a simple answer to the second problem, we have first to introduce the
relevant mathematical concept, namely that of twisted dimensions \cite{PO2},
which are the basic building blocks of the fusion rules of permutation orbifolds.

First, for a rational number \( r=\frac{k}{n} \) in reduced form, i.e. with
positive denominator \( n>0 \) and with \( \gcd (k,n)=1 \), we define the
matrix \( \Lambda (r) \), with matrix elements 
\begin{equation}
\label{Lmatrix}
\Lambda _{pq}(r)=T_{pp}^{-r}M_{pq}T_{qq}^{-r^{*}}
\end{equation}
 where \( M_{pq} \) denotes the matrix elements of the modular transformation
\( M=\left( \begin{array}{cc}
k & y\\
n & x
\end{array}\right)  \), with \( x,y \) any integer solution of the equation \( kx-ny=1 \) - such
solutions always exist because \( \gcd (k,n)=1 \) -, and \( r^{*}=\frac{x}{n} \).
It may be shown that the matrix \( \Lambda (r) \) is well-defined ( after choosing
a definite branch of the logarithm ), and that it is periodic in \( r \), i.e.
\( \Lambda (r+1)=\Lambda (r) \). Notice that \( \Lambda (0)=S \), and more
generally
\[
\Lambda \left( \frac{1}{n}\right) =T^{-\frac{1}{n}}S^{-1}T^{-n}ST^{-\frac{1}{n}}\]

We can now define the twisted dimensions as 
\begin{equation}
\label{tdim}
\mathcal{D}_{g}\left( \begin{array}{ccc}
p_{1} & \ldots  & p_{n}\\
r_{1} & \ldots  & r_{n}
\end{array}\right) =\sum _{q}S_{0q}^{2-2g}\prod _{i=1}^{n}\frac{\Lambda _{qp_{i}}(r_{i})}{S_{0q}}
\end{equation}
 for a non-negative integer \( g \) - the genus -, a sequence of primaries
\( p_{1},\ldots ,p_{n} \) and a sequence of rationals \( r_{1},\ldots ,r_{n} \)
(the characteristics). In case all the characteristics are zero, we have 
\begin{equation}
\label{tverlinde}
D_{g}\left( \begin{array}{ccc}
p_{1} & \ldots  & p_{n}\\
0 & \ldots  & 0
\end{array}\right) =\dim \mathcal{V}_{g}\left( p_{1},\ldots ,p_{n}\right) 
\end{equation}
by Verlinde's formula

After these preliminaries, let's present the solution of the second problem.
According to what has been said above, there is a Verlinde-like formula for
the trace in case the mapping class \( \gamma \in M_{g,0} \) has a fixed point
in its action on the Teichmuller-space \( \mathcal{X}_{g} \). Such a fixed
point \( \tau  \) corresponds to a closed genus \( g \) surface \( \mathcal{S} \)
with a non-trivial automorphism group, and the mapping class \( \gamma  \)
lifts to an automorphism \( \hat{\gamma }\in Aut\left( \mathcal{S}\right)  \)
of finite order \( N \) - actually, one may identify \( Aut\left( \mathcal{S}\right)  \)
with the stabilizer of \( \tau  \) in \( M_{g,0} \) \cite{Hain}. Dividing
out \( \mathcal{S} \) by the action of \( \hat{\gamma } \) we get a new surface
\( \mathcal{S}/\hat{\gamma } \) of genus \( g^{*} \). Of all the orbits of
\( \hat{\gamma } \) on \( \mathcal{S} \), only a finite number have non-trivial
stabilizer subgroups, whose orders we denote by \( n_{1},\ldots ,n_{r} \) -
these orbits correspond to the ramification points of the holomorphic covering
map \( \pi :\mathcal{S}\rightarrow \mathcal{S}/\hat{\gamma } \). In particular,
each \( n_{i} \) divides \( N \). All these quantities are related by the
Riemann-Hurwitz formula 
\begin{equation}
\label{RH}
2g-2=N\left( 2g^{*}-2+\sum _{i=1}^{r}\left( 1-\frac{1}{n_{i}}\right) \right) 
\end{equation}
 The action of \( \hat{\gamma } \) near the \( i \)-th branch point is given
by 
\[
\hat{\gamma }\, :\, z\mapsto \exp \left( 2\pi i\frac{k_{i}}{n_{i}}\right) z\]
for some integer \( 0\leq k_{i}<n_{i} \) coprime to \( n_{i} \), where \( z \)
is a local coordinate. Thus the ratio \( \frac{k_{i}}{n_{i}} \) gives the monodromy
of the covering map \( \pi :\mathcal{S}\rightarrow \mathcal{S}/\hat{\gamma } \)
around the corresponding branch point. 

We are now in position to give a closed expression for the traces. The claim
is that the trace of \( \gamma  \) on \( \mathcal{V}_{g} \) is given by 
\begin{equation}
\label{traces}
Tr(\gamma )=\mathcal{D}_{g^{*}}\left( \begin{array}{ccc}
0 & \ldots  & 0\\
\frac{k_{1}}{n_{1}} & \ldots  & \frac{k_{r}}{n_{r}}
\end{array}\right) 
\end{equation}

Let's illustrate the above results for ordinary modular transformations at genus
1. In this case Teichmuller-space is nothing but the complex upper half-plane
\( \mathcal{X}_{1}=\left\{ \tau \in \mathbb {C}\, |\, Im(\tau )>0\right\}  \),
and the mapping class group is \( M_{1,0}=SL(2,\mathbb {Z}) \), acting on \( \mathcal{X}_{1} \)
via \footnote{%
Why it is \( SL(2,\mathbb {Z}) \) rather than \( PSL(2,\mathbb {Z}) \) is
explained in \cite{Hain}
} 
\[
\left( \begin{array}{cc}
a & b\\
c & d
\end{array}\right) :\tau \mapsto \frac{a\tau +b}{c\tau +d}\]

Note that \( S^{2}=\left( \begin{array}{cc}
-1 & 0\\
0 & -1
\end{array}\right)  \) acts trivially on \( \mathcal{X}_{1} \), i.e. each \( \tau \in \mathcal{X}_{1} \)
is a fixed point of it. The lift of \( S^{2} \) is \( z\mapsto -z \), which
is clearly an involutive automorphism with fixed points \( \left\{ 1,\frac{1}{2},\frac{\tau }{2},\frac{1+\tau }{2}\right\}  \).
According to Eq.(\ref{traces}) we should have 
\begin{equation}
\label{trs2}
Tr\left( S^{2}\right) =\mathcal{D}_{0}\left( \begin{array}{cccc}
0 & 0 & 0 & 0\\
\frac{1}{2} & \frac{1}{2} & \frac{1}{2} & \frac{1}{2}
\end{array}\right) 
\end{equation}

The expression on the rhs. of Eq.(\ref{trs2}) may be rewritten as \( \sum _{p}\nu _{p}^{2} \),
where \( \nu _{p} \) denotes the Frobenius-Schur indicator of the primary \( p \)
- which is \( +1 \) for real, \( -1 \) for pseudo-real, and 0 for complex
primaries \cite{FS}-, and the equality of this last expression with the trace
of \( S^{2} \) follows at once from \( \left( S^{2}\right) _{pp}=\nu _{p}^{2} \)
. While we get nothing new, we have an instance where the trace formula may
be derived by other means.

The other mapping classes of finite order are 

\begin{enumerate}
\item \( S=\left( \begin{array}{cc}
0 & -1\\
1 & 0
\end{array}\right)  \), whose fixed point is \( \tau =i \);
\item \( ST=\left( \begin{array}{cc}
0 & -1\\
1 & 1
\end{array}\right)  \), whose fixed point is \( \tau =\exp \left( \frac{2\pi i}{3}\right)  \);
\item \( \left( ST\right) ^{2}=\left( \begin{array}{cc}
-1 & -1\\
1 & 0
\end{array}\right)  \), with fixed point \( \tau =\exp \left( \frac{2\pi i}{3}\right)  \);
\end{enumerate}
and the inverses of the above, which won't give anything new since \( Tr\left( X^{-1}\right) =\overline{Tr(X)} \)
by the unitarity of the modular representation.

We have to lift the above mapping classes to automorphisms of their fixed points.
Simple considerations show that the lifts are 

\begin{enumerate}
\item \( z\mapsto iz \), of order \( N=4 \).
\item \( z\mapsto \exp \left( \frac{\pi i}{3}\right) z \), of order \( N=6 \).
\item \( z\mapsto \exp \left( \frac{2\pi i}{3}\right) z \), of order \( N=3 \). 
\end{enumerate}
It is straightforward to enumerate the fixed points of the above transformations
(acting on the corresponding tori), and to deduce from this the data \( \left( g^{*};k_{1}/n_{1},\ldots ,k_{r}/n_{r}\right)  \).
The results are summarised in Table 1, where we have expressed the traces in
terms of twisted dimensions. Should we express the relevant twisted dimension
in terms of modular matrix elements, we would get Eq.(\ref{traceS}) for the
trace of \( S \) (and similar results in the other cases). Equating the resulting
expressions with the obvious one for the corresponding modular transformation,
we get the list of all nontrivial trace identities that should hold in any consistent
RCFT.
\begin{table}
\vspace{0.3cm}
{\centering \begin{tabular}{|c|c|c|c|}
\hline 
&
&
&
\\
mapping class&
\( S \)&
\( ST \)&
\( \left( ST\right) ^{2} \)\\
&
&
&
\\
\hline 
\hline 
&
&
&
\\
fixed point &
\( \tau =i \)&
\( \tau =\exp \left( \frac{2\pi i}{3}\right)  \)&
\( \tau =\exp \left( \frac{2\pi i}{3}\right)  \)\\
&
&
&
\\
\hline 
&
&
&
\\
lift&
\( z\mapsto iz \)&
\( z\mapsto \exp \left( \frac{\pi i}{3}\right) z \)&
\( z\mapsto \exp \left( \frac{2\pi i}{3}\right) z \)\\
&
&
&
\\
\hline 
\( N \)&
4&
6&
3\\
\hline 
&
&
&
\\
signature&
\( \left( 0;\frac{1}{4},\frac{1}{4},\frac{1}{2}\right)  \)&
\( \left( 0;\frac{1}{6},\frac{1}{3},\frac{1}{2}\right)  \)&
\( \left( 0;\frac{1}{3},\frac{1}{3},\frac{1}{3}\right)  \)\\
&
&
&
\\
\hline 
&
&
&
\\
trace&
\( \mathcal{D}_{0}\left( \begin{array}{ccc}
0 & 0 & 0\\
\frac{1}{4} & \frac{1}{4} & \frac{1}{2}
\end{array}\right)  \) &
\( \mathcal{D}_{0}\left( \begin{array}{ccc}
0 & 0 & 0\\
\frac{1}{6} & \frac{1}{3} & \frac{1}{2}
\end{array}\right)  \)&
\( \mathcal{D}_{0}\left( \begin{array}{ccc}
0 & 0 & 0\\
\frac{1}{3} & \frac{1}{3} & \frac{1}{3}
\end{array}\right)  \)\\
&
&
&
\\
\hline 
\end{tabular}\par}\vspace{0.3cm}

\caption{Trace formulae for \protect\( g=1\protect \).}
\end{table}
 \\

Let's now turn to the origin of the trace formulae. For this we need to recall
that to any RCFT there corresponds a 3D Topological Field Theory \cite{Witten},
and that this later assigns a complex number \( Z(M) \) - the partition function
- to each 3 dimensional closed manifold, such that \( Z(M_{1})=Z(M_{2}) \)
whenever \( M_{1} \) and \( M_{2} \) are homeomorphic. The partition function
of a given 3-manifold can be determined in principle from the knowledge of the
modular matrix elements of the RCFT via surgery, although this is by no means
a simple task in general. The point is that for some classes of 3-manifolds
one can give closed expressions for the partition function.

One such class is that of the so-called fibred manifolds \cite{CZ}. These are
obtained by the following procedure: take a closed surface \( \mathcal{S} \)
of genus \( g \), and form the product \( \mathcal{S}\times [0,1] \). The
resulting 3-manifold is not closed, as it has two boundary components, each
homeomorphic to \( \mathcal{S} \), so we have to glue together these boundary
components. When doing so, one has the freedom to identify the boundary components
via a self-homeomorphism \( \gamma :\mathcal{S}\rightarrow \mathcal{S} \).
As it turns out, the topological equivalence class of the resulting closed 3-manifold
does not depend on the actual choice of \( \gamma  \), but only on its mapping
class \( \left[ \gamma \right]  \). This implies that for each \( \gamma \in M_{g,0} \)
we get a closed 3-manifold \( F_{\gamma } \), well-defined up to topological
equivalence. It might not come as a big surprise that the partition function
of \( F_{\gamma } \) is just the trace of the operator representing \( \gamma  \)
on the space of genus \( g \) characters, i.e. 
\begin{equation}
\label{fiber}
Z(F_{\gamma })=Tr(\gamma )
\end{equation}

There is another class of 3-manifolds for which we know the partition function,
the Seifert-manifolds. They may be constructed according to the following recipe:
consider a closed (oriented) surface \( \mathcal{S} \) of genus \( g \), and
cut out \( n \) non-overlapping disks to obtain a surface \( \mathcal{S}' \).
The product \( \mathcal{S}'\times S^{1} \) is not closed, its boundary consisting
of \( n \) disjoint 2-tori. To get a closed 3-manifold, one has to paste in
solid tori to these boundary components, and in doing so, one has to glue the
\( i \)-th boundary component to the boundary of the corresponding solid torus
by means of a modular transformation \( M_{i} \). As it turns out, the resulting
Seifert-manifold \( S(g;r_{1},\ldots ,r_{n}) \) is characterised, besides the
genus \( g \), by the sequence \( r_{i}=M_{i}(i\infty ) \) of rationals modulo
integers, i.e. the images of the cusp at infinity. The partition function of
a Seifert-manifold (\cite{Freed},\cite{Rozansky}) in terms of twisted dimensions
reads 
\begin{equation}
\label{seifert}
Z\left[ S(g;r_{1},\ldots ,r_{n})\right] =\mathcal{D}_{g}\left( \begin{array}{ccc}
0 & \ldots  & 0\\
r_{1} & \ldots  & r_{n}
\end{array}\right) 
\end{equation}

We are nearly done, all that remains is to notice that the class of Seifert-manifolds
and that of fibred manifolds overlap. Actually, a fibred manifold \( F_{\gamma } \)
is a Seifert-manifold exactly when the mapping class \( \gamma  \) has a fixed
point in its action on Teichmuller-space, i.e. when it has finite order \cite{Hempel}.
So for such \( \gamma \in M_{g,0} \), \( F_{\gamma } \) is homeomorphic to
\( S(g^{*};r_{1},\ldots ,r_{n}) \) for some signature \( \left( g^{*};r_{1},\ldots ,r_{n}\right)  \),
consequently their partition functions coincide, i.e. 
\[
Tr(\gamma )=\mathcal{D}_{g^{*}}\left( \begin{array}{ccc}
0 & \ldots  & 0\\
r_{1} & \ldots  & r_{n}
\end{array}\right) \]
 It remains to determine the signature \( \left( g^{*};r_{1},\ldots ,r_{n}\right)  \)
corresponding to a given mapping class \( \gamma  \) of finite order. The detailed
analysis \cite{Hempel} leads to the result presented earlier: for \( \gamma \in M_{g,0} \)
of finite order, \( F_{\gamma } \) is homeomorphic to a Seifert-manifold whose
signature is determined by the monodromy of the covering map \( \pi :\mathcal{S}\rightarrow \mathcal{S}/\hat{\gamma } \). 

One may look at the above results from a different perspective. Let \( g>1 \),
and let's take some point \( \tau \in \mathcal{X}_{g} \) in Teichmuller-space,
and suppose that the corresponding Riemann-surface \( \mathcal{S} \) has a
non-trivial automorphism group \( \mathcal{A} \) (which is known to be finite).
To each element of \( \mathcal{A} \) corresponds a mapping class from \( M_{g,0} \),
and this correspondence is a homomorphism, consequently \( \mathcal{A} \) is
represented on the space of genus \( g \) characters. But it follows from the
results above that this representation is completely determined by the modular
representation on the space of genus 1 characters, because we know the traces
of all the representation operators. In other words, the space of genus \( g \)
characters affords representations of the automorphism groups of all genus \( g \)
closed surfaces, and these representations are completely determined by genus
1 data. This puts severe arithmetic restrictions on the allowed values of twisted
dimensions, and hence on modular matrix elements, which might prove useful in
classification attempts. 

As an example, let's consider the Klein quartic, i.e. the surface of genus 3
with the maximum number (=168) of automorphisms allowed by Hurwitz's theorem.
The automorphism group \( \mathcal{A} \) is isomorphic to \( SL(3,2) \), and
it contains the following nontrivial elements  

\vspace{0.3cm}
{\centering \begin{tabular}{|c|c|c|c|}
\hline 
order&
number&
\#fixed points&
trace\\
\hline 
\hline 
&
&
&
\\
2&
21&
4&
\( \mathcal{D}_{1}\left( \begin{array}{cccc}
0 & 0 & 0 & 0\\
\frac{1}{2} & \frac{1}{2} & \frac{1}{2} & \frac{1}{2}
\end{array}\right)  \)\\
&
&
&
\\
\hline 
&
&
&
\\
3&
56&
2&
\( \mathcal{D}_{1}\left( \begin{array}{cc}
0 & 0\\
\frac{1}{3} & \frac{2}{3}
\end{array}\right)  \)\\
&
&
&
\\
\hline 
&
&
&
\\
4&
42&
0&
\( \mathcal{D}_{1}\left( \begin{array}{cc}
0 & 0\\
\frac{1}{2} & \frac{1}{2}
\end{array}\right)  \)\\
&
&
&
\\
\hline 
&
&
&
\\
7&
48&
3&
\( \mathcal{D}_{0}\left( \begin{array}{ccc}
0 & 0 & 0\\
\frac{1}{7} & \frac{2}{7} & \frac{4}{7}
\end{array}\right)  \) \\
&
&
&
\\
\hline 
\end{tabular}\par}
\vspace{0.3cm}

The knowledge of the representations of \( \mathcal{A} \) allows us to deduce
e.g. the conditions 
\begin{eqnarray*}
\mathcal{D}_{3}-\mathcal{D}_{0}\left( \begin{array}{ccc}
0 & 0 & 0\\
\frac{1}{7} & \frac{2}{7} & \frac{4}{7}
\end{array}\right)  & \in  & 7\mathbb {Z}_{+}\\
\mathcal{D}_{3}-\mathcal{D}_{1}\left( \begin{array}{cc}
0 & 0\\
\frac{1}{3} & \frac{2}{3}
\end{array}\right)  & \in  & 3\mathbb {Z}_{+}\\
\mathcal{D}_{3}-\mathcal{D}_{1}\left( \begin{array}{cccc}
0 & 0 & 0 & 0\\
\frac{1}{2} & \frac{1}{2} & \frac{1}{2} & \frac{1}{2}
\end{array}\right)  & \in  & 4\mathbb {Z}_{+}
\end{eqnarray*}
with \( \mathcal{D}_{3} \) denoting the number of genus 3 characters. These
nontrivial congruences should hold in any consistent RCFT. For example, in the
case of the Ising model we find that the representation of \( \mathcal{A} \)
on the space of genus 3 characters contains 4 copies of the trivial representation,
4 copies of the 6 dimensional irrep, and one copy of the 8 dimensional irrep.

In summary, we have found that traces of mapping classes of finite order are
determined by the modular representation through Verlinde-like formulae, which
in the \( g=1 \) case lead to interesting trace identities for modular matrix
elements. The origin of these trace formulae is the overlap between the classes
of Seifert- and of fibred 3-manifolds. The representation of the automorphism
groups of surfaces on the space of characters is also determined by the genus
1 data, and this leads to interesting arithmetic restrictions on the allowed
modular representations in consistent RCFTs. 

\emph{Thanks for the hospitality of ESI, where this work was completed. Research
supported by grant OTKA T32453.}

\end{document}